\documentclass[preprint,authoryear,11pt]{article}      
\usepackage[a4paper,top=3cm,left=2.3cm,right=2.2cm,bottom=3cm]{geometry}

\usepackage{graphicx}
\usepackage{epstopdf}
\usepackage{dcolumn}
\usepackage{bm}
\usepackage{color}
\usepackage{authblk}
\usepackage{amssymb, amsmath, bm, textcomp}

\usepackage[T1]{fontenc} 
\usepackage{float}       
\usepackage{mathrsfs}

\usepackage{setspace}    
\usepackage{color}
\usepackage[table]{xcolor}
\usepackage{subfigure}
\usepackage{caption}

\newcommand{\be}{\begin{equation}}
\newcommand{\ee}{\end{equation}}
\newcommand{\bea}{\begin{eqnarray}}
\newcommand{\eea}{\end{eqnarray}}

\begin{document}



\title{\LARGE{\textbf{Low pH, high salinity: too much for Microbial Fuel Cells?}}}



\author[1]{Nicole Jannelli}
\author[2]{Rosa Anna Nastro}
\author[3]{Viviana Cigolotti} 

\author[2]{Mariagiovanna Minutillo}
\author[4]{Giacomo Falcucci}

\affil[1]{A.T.E.N.A. scarl \\
Centro Direzionale - Isola C4, 80143 Naples, Italy}
\affil[2]{Dept. of Engineering - University of Naples  ``Parthenope'', \\
 Centro Direzionale - Isola C4, 80143 Naples, Italy}
 \affil[3]{ENEA - Portici Research Center \\
P.le Enrico Fermi 1 - 80055 Portici (Na)} 
\affil[4]{Dept. of Enterprise Engineering ``Mario Lucertini'' - University of Rome  ``Tor Vergata'', \\
 Via del Politecnico 1, 00100 Rome, Italy \\

\vspace{1.5cm}

\textbf{To cite this paper:} Jannelli, N., Nastro, R.A., Cigolotti, V., Minutillo, M. and Falcucci G. Low pH, high salinity: Too much for microbial fuel cells?, \textit{Applied Energy}, Available online 3 August 2016, ISSN 0306-2619, http://dx.doi.org/10.1016/j.apenergy.2016.07.079.  }

\vspace{1.5cm}
{

 }
%
%
%
\date{Sept., 1st, 2016}
 \maketitle

\begin{abstract}

Twelve single chambered, air-cathode Tubular Microbial Fuel Cells (TMFCs) have been filled up with fruit and vegetable residues. The anodes were realized by means of a carbon fiber brush, while the cathodes were realized through a graphite-based porous ceramic disk with Nafion membranes (117 Dupont). 
The performances in terms of polarization curves and power production were assessed according to different operating conditions: percentage of solid substrate water dilution, adoption of freshwater and a 35mg/L NaCl water solution and, finally, the effect of an initial potentiostatic growth.

All TMFCs operated at low pH (pH$=3.0 \pm 0.5$), as no pH amendment was carried out. 
Despite the harsh environmental conditions, our TMFCs showed a Power Density (PD) ranging from 20 to 55~mW/m$^2 \cdot$kg$_{\text{waste}}$ and a maximum CD of  20~mA/m$^2 \cdot$kg$_{\text{waste}}$, referred to the cathodic surface. \\
COD removal after a $28-$day period was about $45 \%$.

The remarkably low pH values as well as the fouling of Nafion membrane very likely limited TMFC performances. However, a scale-up estimation of our reactors provides interesting values in terms of power production, compared to actual anaerobic digestion plants. 
These results encourage further studies to characterize the graphite-based porous ceramic cathodes and to optimize the global TMFC performances, as they may provide a valid and sustainable alternative to anaerobic digestion technologies.

\end{abstract}

\textbf{Keywords}: \\
microbial fuel cell; waste to energy; power generation; direct conversion of solid organic waste \\






\section{Introduction}

The quest for sustainable and efficient energy sources has been the object of great research efforts in the last decades, \cite{An_2011,Srirangan_2012}. Along with the increasing need of a more sustainable waste management, new approaches able to combine waste treatment with energy recovery can contribute to the establishment of a green economy \cite{Karagiannidis_2009,Koufodimos_2002,Falcucci_CCHP_2014,Nastro_et_al_2016}. Generally, waste-to-energy technologies based on incineration, gasification or anaerobic digestion are recommended for large-scale plants, to produce thermal energy or to convert waste in biogas, syngas and other secondary fuels, \cite{Karagiannidis_2009, Koufodimos_2002,Pandey_2016}. Nevertheless, the large part of these processes do not require the selection of incoming waste, with a consequent, unavoidable production of pollutants, toxic leachate, ash and greenhouse gases, \cite{Pavlas_2009}. The treatment of such pollutants makes the management and control of these processes complex and expensive, \cite{Cormos_2014,Dong_2014}. \\
In this context, bio-reactors based on natural processes like fermentation and/or methanogenesis, can play an important role towards the establishment of a more sustainable waste management. Microbial Fuel Cells (MFCs) are based on the ability of microorganisms to use inorganic compounds as electronic acceptor to obtain electric power directly from microbe metabolism, without any combustion. Recent researches confirm the potential of such systems in turning ``waste'' into a ``resource'', by treating different types of substrates and even recovering by-products with a non-negligible economic potential, \cite{Nastro_et_al_2016}. \\
As biomass-based systems, MFCs (like other Bioelectrochemical Systems) are considered carbon neutral: the bio-transformation of organic matter into chemicals through microbial metabolism, in fact, prevents the primary production of  CO$_2$ emissions. Moreover, MFCs do not involve CH$_4$ production and combustion, as opposed to traditional anaerobic digestion plants. Thus, this newborn technology is characterized by remarkable features: the direct electrical power production, the conversion of the chemical energy contained in any form of biomass \cite{Mohan_2014}, low environmental impact, low operating temperatures and simple architectures \cite{Du_2007}.  All these characteristics, together with the environmental advantages ingrained with this technology are supposed to largely overcome the costs for MFC development and implementation, \cite{Nastro_et_al_2016}. 

However, even though a wide number of papers deals with the set-up and the study of MFCs fed with wastewaters, few attempts to apply this innovative technology to solid organic waste treatment have been carried out. In 2011, Mohan \& Chandrasekhar studied the operational factors affecting the performances of MFCs fed with canteen food waste, focusing on electrodes distance and feedstock pH, \cite{Mohan_2011}. After this seminal work, other researchers started working on the application of MFCs to the Organic Fraction of Municipal Solid Waste (OFMSW) using different approaches, but all confirming the effectiveness of MFC technology as tool for energy recovery and organic load removal from OFMSW, \cite{Nastro_et_al_2016}.
MFCs, in fact, represent a valid alternative to achieve small-scale, distributed and efficient conversion of organic waste into electricity \cite{Logan_2006}, even in developing Countries, where solid waste is often dumped rather than landfilled. \\
Nevertheless, the industrial implementation of this technology is still challenging, \cite{Di_Maria_2015,Di_Palma_2015}. The use of solid or liquid food waste in MFCs generally leads to anodic environmental conditions highly unfavorable for bioelectricity production, due to low pH and high salinity and, as a consequence, high ionic strength, \cite{Mohan_2014,Li_2016,Karthikeyan_2016,Choi_2015}. \\
Thus, the scale-up of reactors, necessary for industrial power applications, is limited by several issues, so that further research is required to improve the reactor geometry, the economic feasibility and the response of the system to unfavorable conditions. In our previous works, we assessed the possible adoption of MFC technology for the treatment of OFMSW, highlighting the importance of appropriate reactor layout and electrode design, in order to minimize the internal resistance, \cite{Frattini_2016,Falcucci_2013_MFC,Nastro_EFC_2015,Nastro_Cleaner_2015}. \\

In this work, we explored the efficiency of tubular single-chambered, air-cathode MFCs (TMFCs) fed by a feedstock composed of vegetable and fruit slurry. All MFCs were provided with Nafion membranes adhering to the 
cathode, which was realized by means of a graphite-based porous ceramic disk: this innovative material was ad-hoc designed and realized in order to be used in scaled-up MFCs, as well. Power production was studied as a function of different parameters, such as solid substrate inoculum, liquid-to-solid ratio, salinity of amending solution and the adoption of a preliminary potentiostatic growth phase. An ad-hoc measurement set-up has been realized to test the MFC reactors in order to prove evidence of their feasibility and reliability in standard and also harsh anodic environment.

\section{Materials \& Methods}

\subsection{TMFC Assembly}

\begin{figure}
\centering
\includegraphics[height=5.5cm]{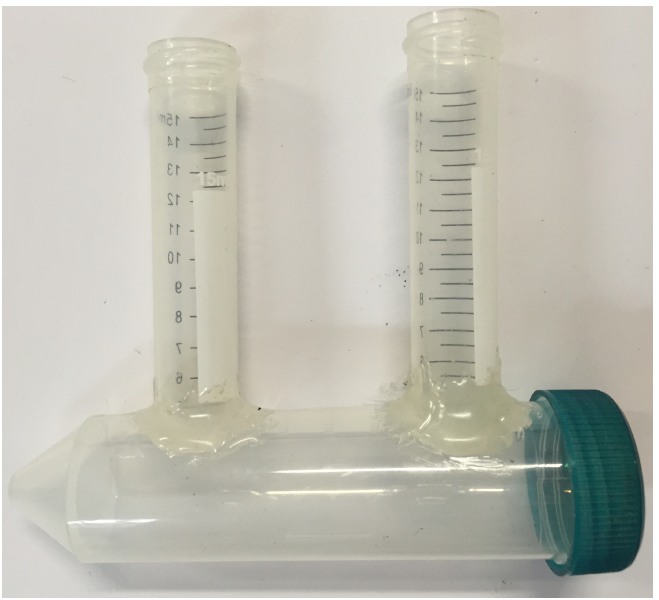} \\
\caption{\label{Fig_1a} Layout of the Tubular MFC, realized by means of one 50 mL and two 20 mL Falcon test tubes.}
\end{figure}

Twelve tubular MFC bioreactors, adapted from \cite{Logan_2007,Kim_2009,Frattini_2016}, were realized by using standard 50~mL polypropylene Falcon test tubes, supplied by BD Corning Inc. (Tewksbury, USA), sterile and suitable for biological cultures. Two 20 mL Falcon tubes were used to sample the organic feedstock and to monitor pH (see Fig. \ref{Fig_1a}).

\begin{figure}
\centering
\subfigure[]{\includegraphics[height=3.5cm]{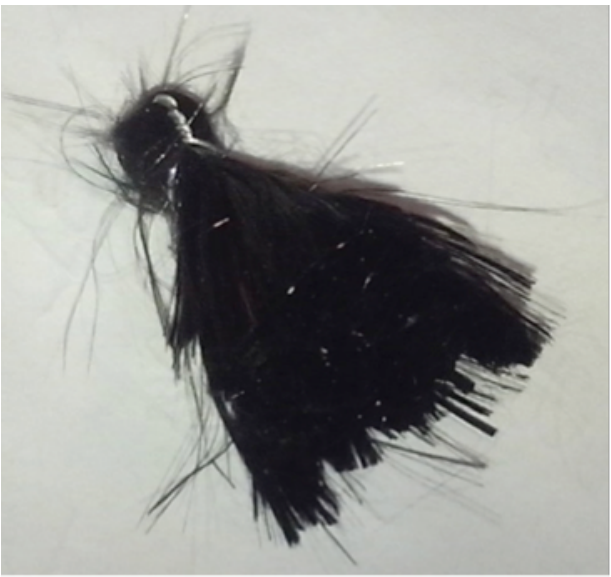} } \qquad
\subfigure[]{\includegraphics[height=3.5cm]{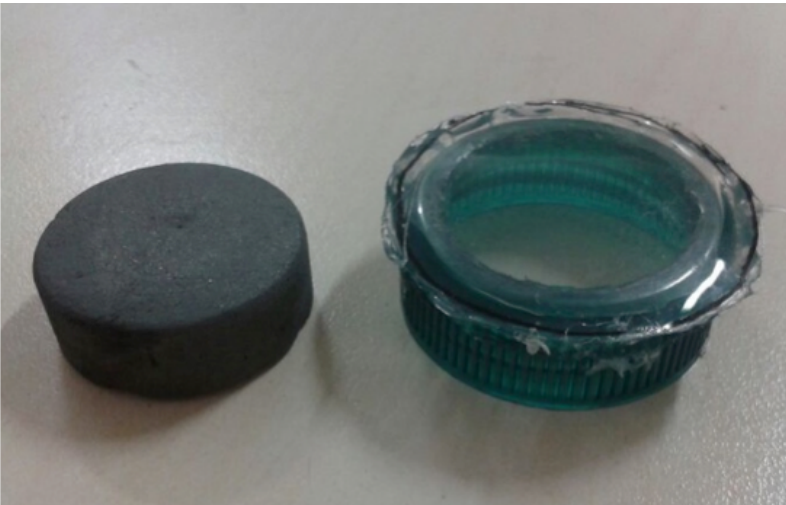}} \\
\caption{\label{Fig:1b} Detail of the Anode brush (a) and Cathode Nafion assembly electrode (b).}
\end{figure}

The electrodes were made by carbon-fiber anode brush, realized with a high strength carbon fiber from FIDIA s.r.l. (Perugia, ITALY) and unpolished stainless steel wire (ASTM A313) with 0.5 mm section, while for the cathode a porous ceramic disk was developed starting from graphite powder type GK 2 Ultra-fine, by AMG Mining AG (Hauzenberg, Germany). The brush anode had an estimated surface area of 0.22 m$^2$/g while the cathode disk had a surface area of 60.75 m$^2$/g, \cite{Logan_2007}, (Figure \ref{Fig:1b}(b)). The electrodes were placed at a distance of $\sim 3$ cm.  A standard Nafion 117 membrane by DuPont Inc. (Richmond, USA, \cite{Nafion}) was used to seal the porous cathode.


\subsection{Data Acquisition System}

\begin{figure}[h!]
\centering
\includegraphics[width=0.73\textwidth]{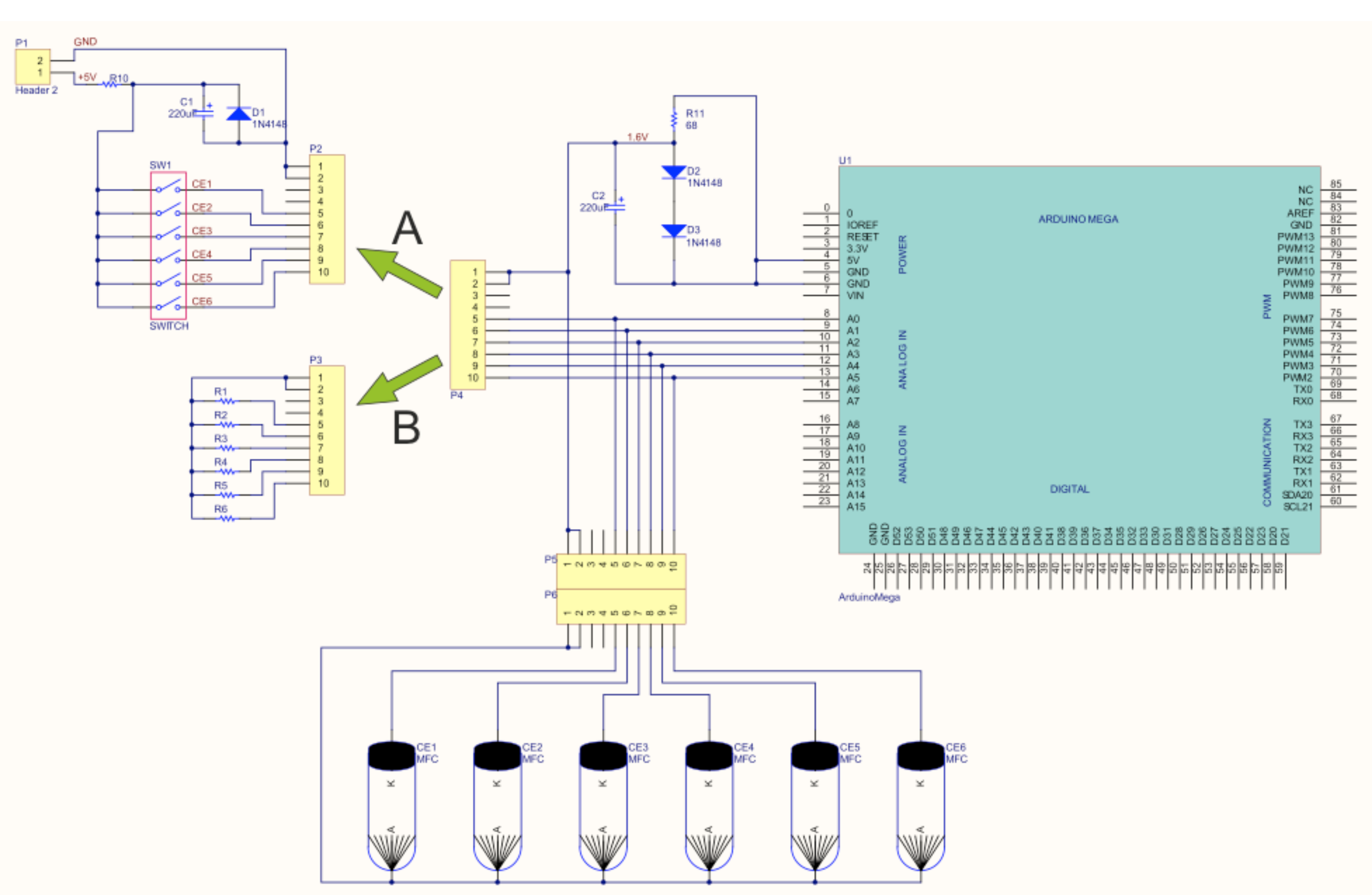}
\caption{\label{Fig_arduino} Electrical Scheme of the acquisition and measurement system.}
\end{figure}

The data collection hardware was based on the Arduino board MEGA 2560, \cite{arduino}, composed by a load array (for polarization curve acquisition) with 6 resistors, ranging from $\sim 10^6$ $\Omega$ to $\sim 10 \ \Omega$. The software for data acquisition was developed with LabVIEW Interface For Arduino, (LIFA) package. Figure \ref{Fig_arduino} sketches the acquisition system and the measurement chain.

\subsection{Experimental Setup}
\label{setup}
Apples, pumpkins, chickpeas and zucchini in a 1:1:1:1 ratio were used to prepare a slurry by mechanically mixing the vegetable residues with water (see Tab \ref{Tab_2} for the details of water composition). \\
In order to verify the effect of salt concentration on TMFC performances, six reactors were filled with a slurry prepared using fresh water; in the other six cells, a NaCl solution (35~mg/L NaCl) was added to improve the slurry conductivity and test TMFC performances in presence of high Na$^+$ concentration, as salts concentration can increase, in leachate, along with the organic matter degradation, \cite{Mohan_2014,Li_2016,Karthikeyan_2016,Choi_2015}. \\ 
The Liquid-to-Solid ratio, and the specific conditions used in each cell are reported in Table \ref{Tab_3}. 
Cells 4, 5 and 10, 11 have been added with spent substrate (10$\%$ of the overall feedstock) from TMFC reactors filled with the same organic substrate and previously operated for a 28-day period. \\
Cells 1, 2, 3, 6, 7, 8 and 9 were filled with 30~g of solid waste and 30~g of water, while cells 4, 5, 10, 11 and 12 were filled by 15~g of solid waste and 45~g of water, to obtain the desired Liquid-to-Solid ratio. 
Finally, cells 3 and 9 were subject to a potentiostatic growth phase for the first 7 days of operation, by imposing an external potential difference between the anode and the cathode of $\Delta V_{ext}= 0.8$ V: this procedure was aimed at verifying the possible effects of narrowing the bacteria consortium to a mixed culture of exoelectrogenic micro-organisms.
\begin{table}
\begin{center}
\begin{tabular}{l c}
\hline
pH 	\; &	6.65 $[-]$ \\
Electrical Conductibility (at 20$^{\circ}$C)  \;  & 420 $\mu$S/cm \\
Fixed residue (at 180$^{\circ}$C)   \;  & 341 mg/L \\
Free CO$_2$ at source    \;  & 125 mg/L \\
Bicarbonates \;  & 254 mg/L \\ 
Potassium \;  & 26.9 mg/L \\
Calcium \;  &   31.5 mg/L  \\
Magnesium \;  &  9.4 mg/L \\
Fluorides \;  & 1.1 mg/L \\
Nitrates  \;  & 7 mg/L \\
\hline
\end{tabular}
\end{center}
\captionof{table}{\small{\label{Tab_2} Chemical composition of the fresh water used to dilute the solid organic substrate.}}
\end{table} 

\begin{table}[ht]
\rowcolors{1}{gray!20}{white}
\begin{center}
\begin{tabular}{c c c c c c}
\hline
\rowcolor{white}
Cell $\#$ \;  	&  Fresh		&		Salt 			& Potentiostatic & Spent substrate 	&	Liquid-to-Solid \\
					&  water 		&		water		& conditioning   &  amendment			&	ratio \\
\hline	
\hline				
1					& 	X				&						&						&								&  1:1 \\
2					& 	X				&						&						&								&  1:1 \\
3					& 	X				&						&			X			&								&  1:1 \\
4					& 	X				&						&						&			X					&  3:1 \\
5					& 	X				&						&						&			X					&  3:1 \\
6					& 	X				&						&						&								&  3:1 \\
7					& 					&	X					&						&								&  1:1 \\
8					& 					&	X					&						&								&  1:1 \\
9					& 					&	X					&			X			&								&  1:1 \\
10					& 					&	X					&						&			X					&  3:1 \\
11					& 					&	X					&						&			X					&  3:1 \\
12					& 					&	X					&						&								&  3:1 \\
\hline
\end{tabular}
\end{center}
\caption{\small{\label{Tab_3} Characteristics of the different realized MFC's.}}
\end{table}

\subsection{Chemical Analyses}

Slurry organic content in terms of COD was measured according to Standard Methods (2012), \cite{Standard_Methods_2012}, at beginning of the experiment and after 28 days in order to assess MFCs efficiency in organic load removal. 
NO$_3^-$ and NO$_2^-$ concentrations were evaluated as well, in order to provide further
information regarding the micro-organism metabolism inside the TMFC reactors.

\begin{table}
\begin{center}
\begin{tabular}{l c c c}
\hline
                                 &  COD           &   NO$_3^-$        &   NO$_2^-$ \\
\hline
Fresh  Substrate       &	48960          &     40         &       9.6  \\
Spent Substrate	   	&	27520          &    180        &      12.2 \\
Mixture                     & 	48320          &      140      &        11.4 \\
\hline
\end{tabular}
\end{center}
\captionof{table}{\small{\label{Tab_COD} Chemical characteristics of the employed feedstocks: all the values are expressed in mg/L.}}
\end{table}

\section{Results \& Discussion}

\subsection{Polarization Behavior}

\begin{figure*}
\centering
\subfigure[11th Day of Operation]
{\includegraphics[angle=-90,width=0.32\textwidth]{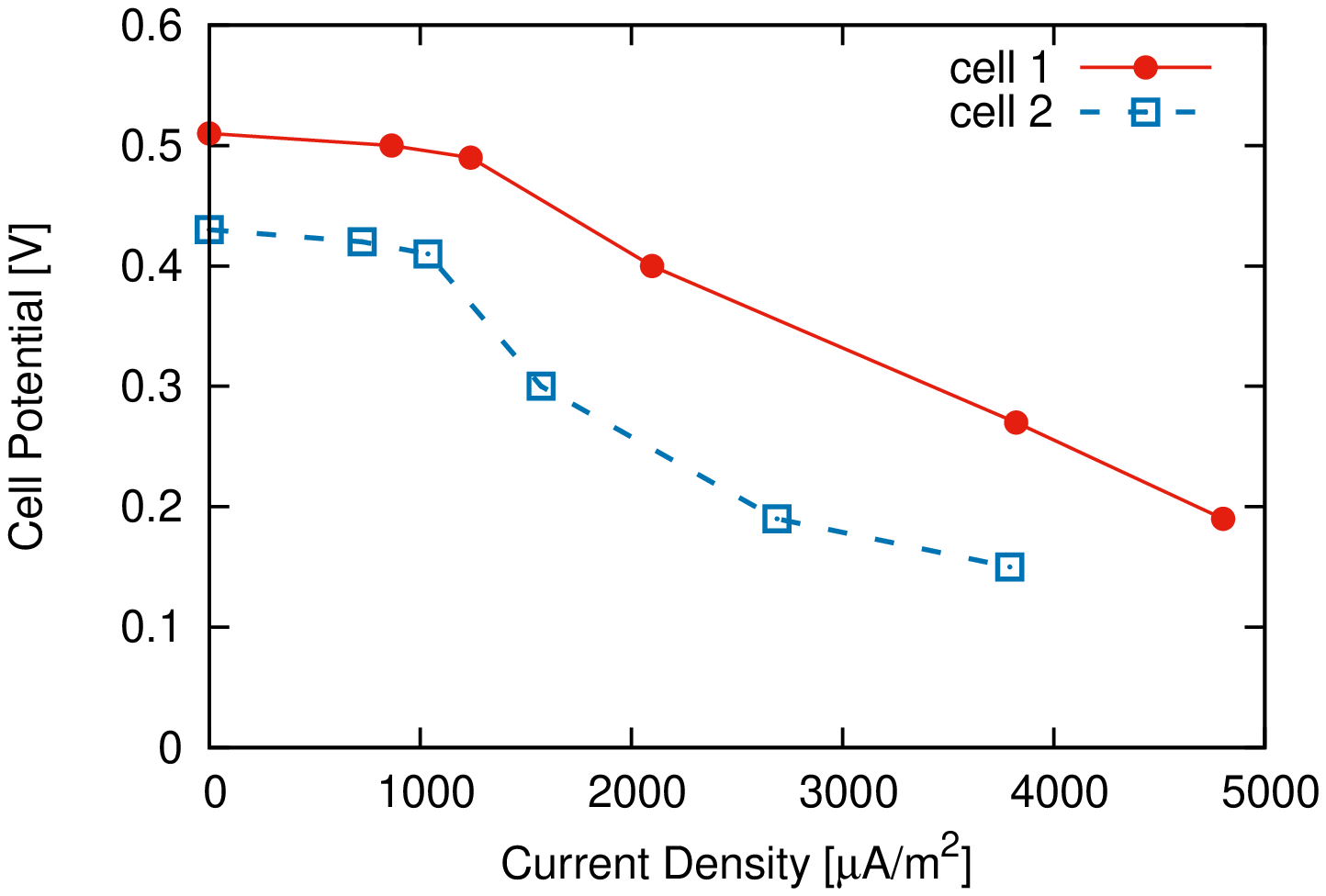} \
\includegraphics[angle=-90,width=0.32\textwidth]{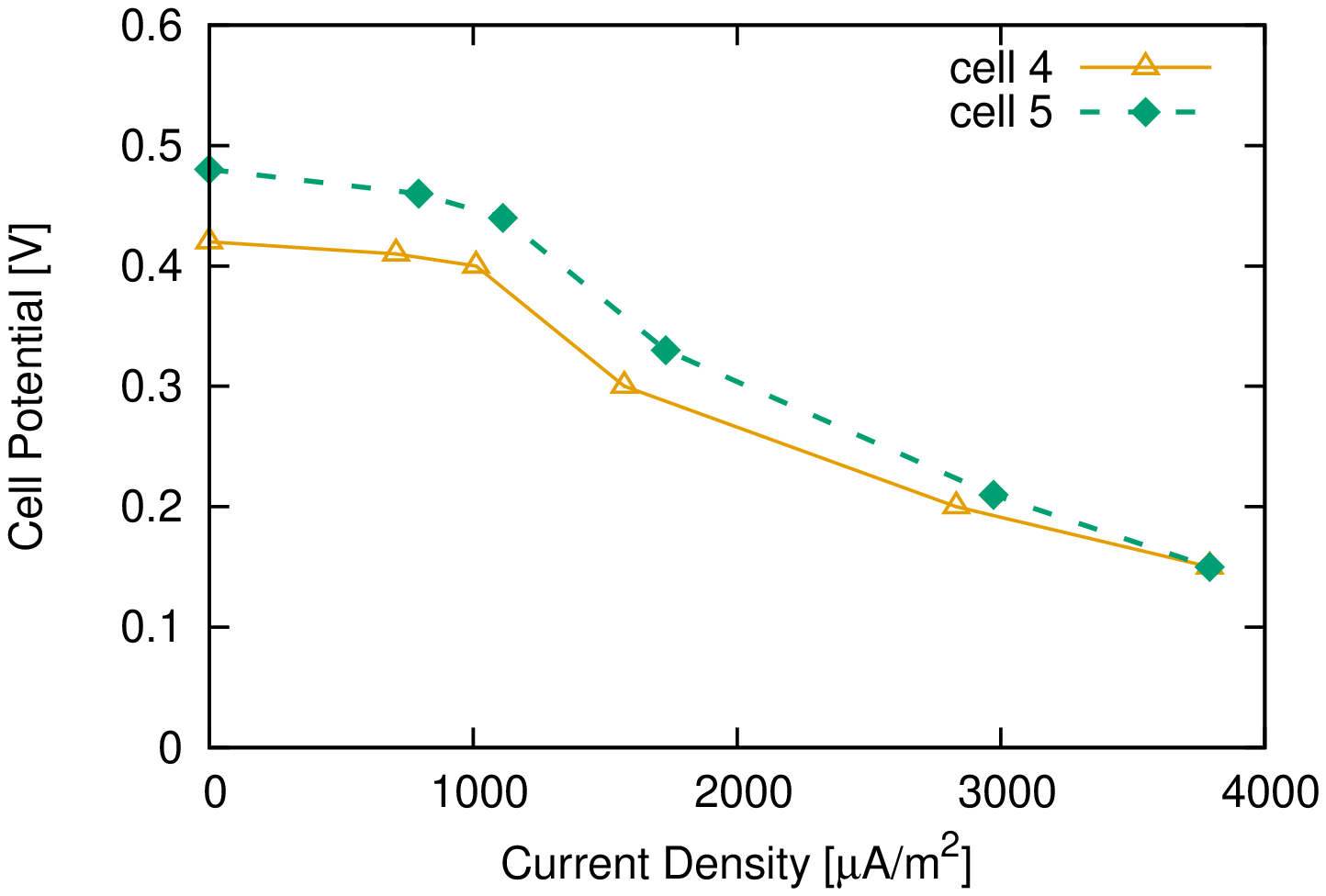} \
\includegraphics[angle=-90,width=0.32\textwidth]{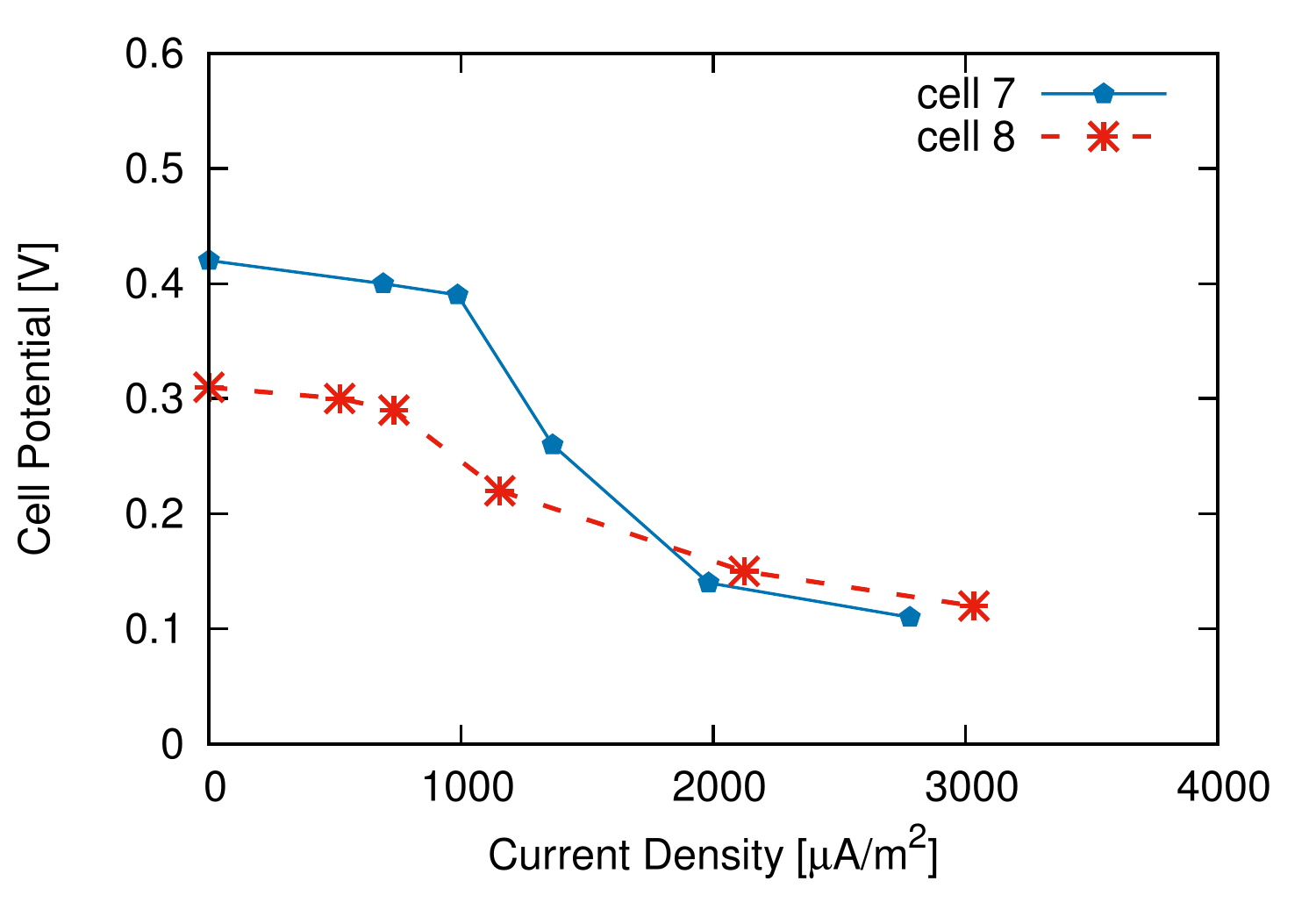} }
\subfigure[21th Day of Operation]
{\includegraphics[angle=-90,width=0.32\textwidth]{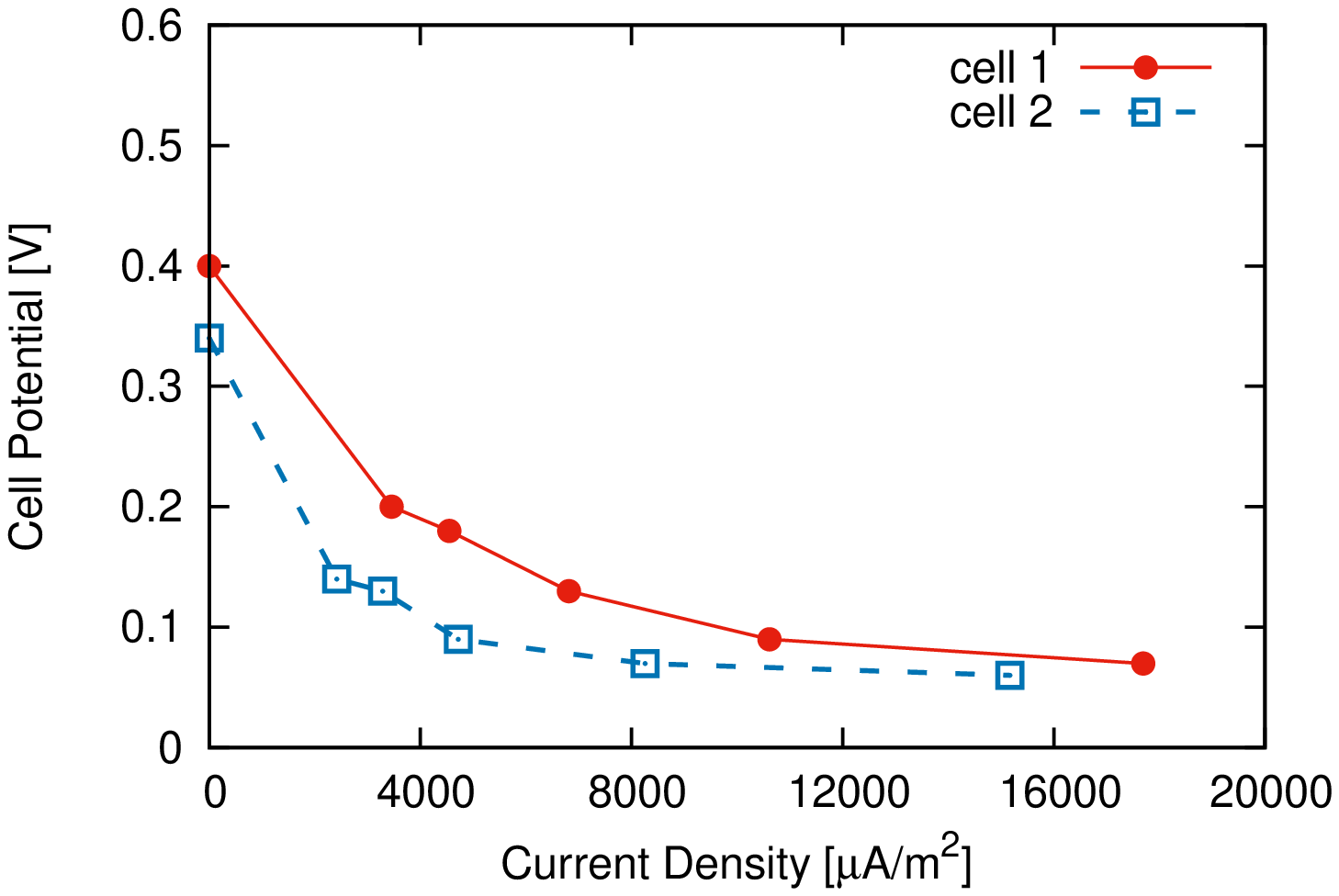} \
\includegraphics[angle=-90,width=0.32\textwidth]{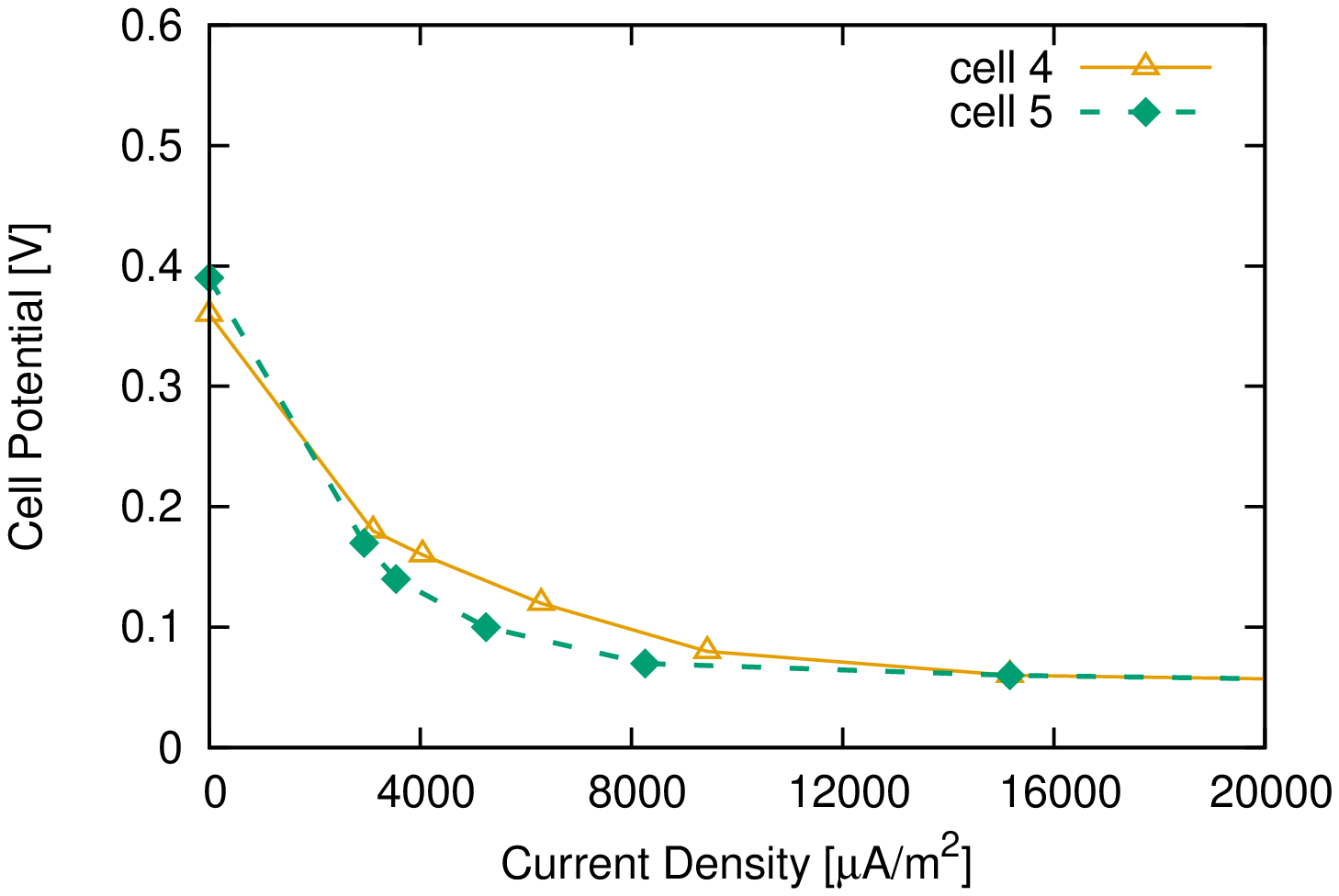} \
\includegraphics[angle=-90,width=0.32\textwidth]{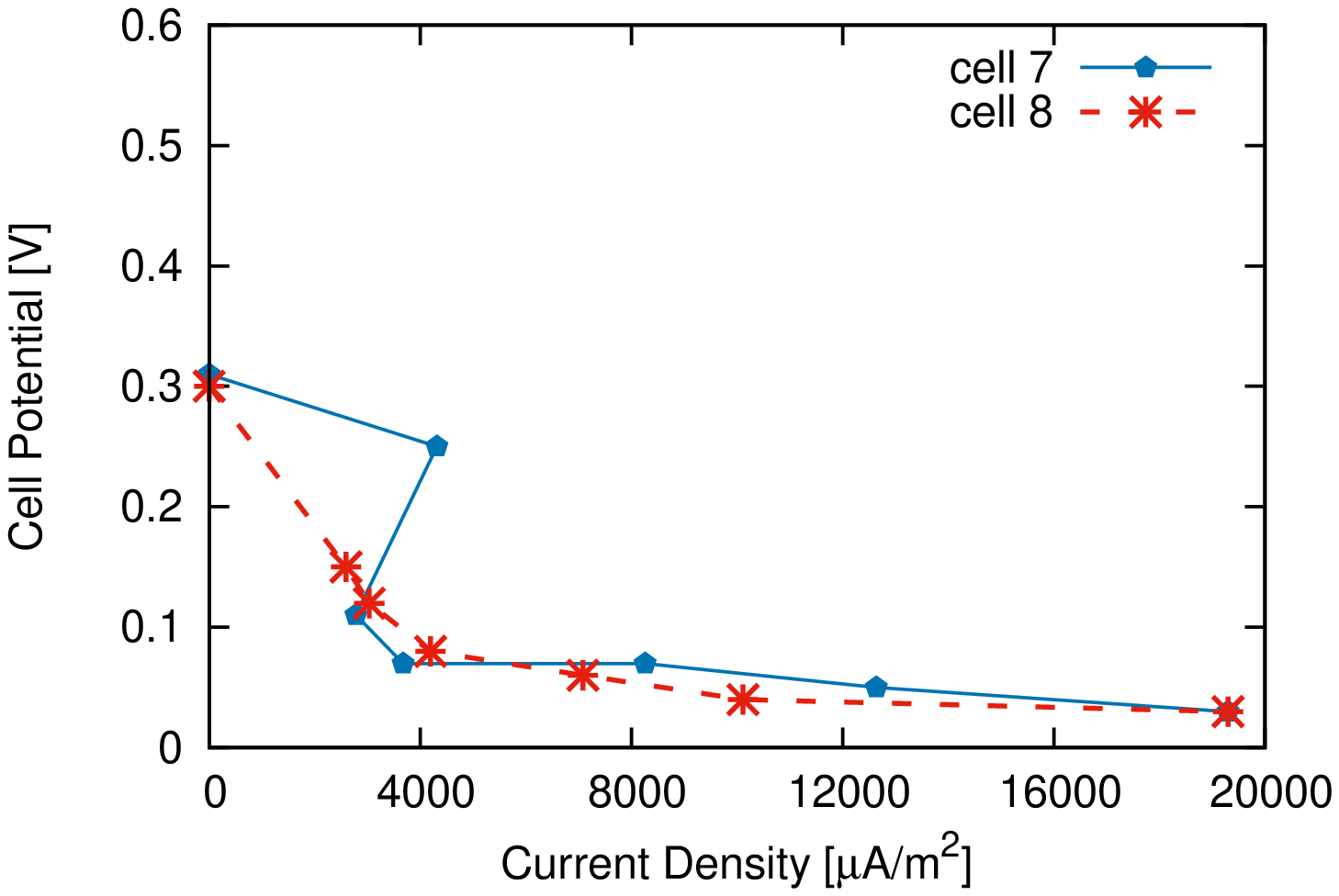} }
\caption{\label{Fig_3} Comparison of the performances of TMFCs, in terms of polarization curves, at different operating days, arranged according to the feedstock characteristics and the operating conditions.}
\end{figure*}

Figure \ref{Fig_3} reports the polarization curves of the different cells, according to the feedsotck concentration and to its salinity; in the Figure, the performance of the cells subject to the initial potentiostatic growth process are reported, as well. \\
According to Fig. \ref{Fig_3}, the cells highlight similar polarization trends, despite the differences in the substrate. 
This can be ascribed to the layout of the cells and gives evidence of the reproducibility of TMFCs performances. \\
Salt-water TMFCs were characterized by a different behavior, compared to fresh-water cells: salty reactors, in fact, showed a lower value of internal resistance (R$_{\text{int}}$, given by the slope of polarization curves) in the first 7 days of the experiment but, during the last week of operation, the internal resistance of these cells became higher than the R$_{\text{int}}$ value of freshwater TMFCs. It is known that an increase in the feedstock conductivity results in a decrease in the cell resistance: in fact, cations concentration increases and, thus, their flow towards the cathode, where they take part to the electrochemical reactions. \\
However, in presence of high cation concentration, Nafion sulfonated groups are rapidly saturated, preventing the protons linkage to the membrane itself, \cite{Chae_et_al_2008}.  This mechanism, known as Nafion \textit{fouling}, together with the biofilm growth, leads to the saturation of the membrane, which eventually prevents cations and protons from taking part to the cathodic reactions: this causes a decrease of cathode functionality, an increase in the cell internal resistance (compared to freshwater reactors) and, as a consequence, a degradation of TMFC performance. \\

\begin{figure}
\centering
	\includegraphics[width=0.75\textwidth]{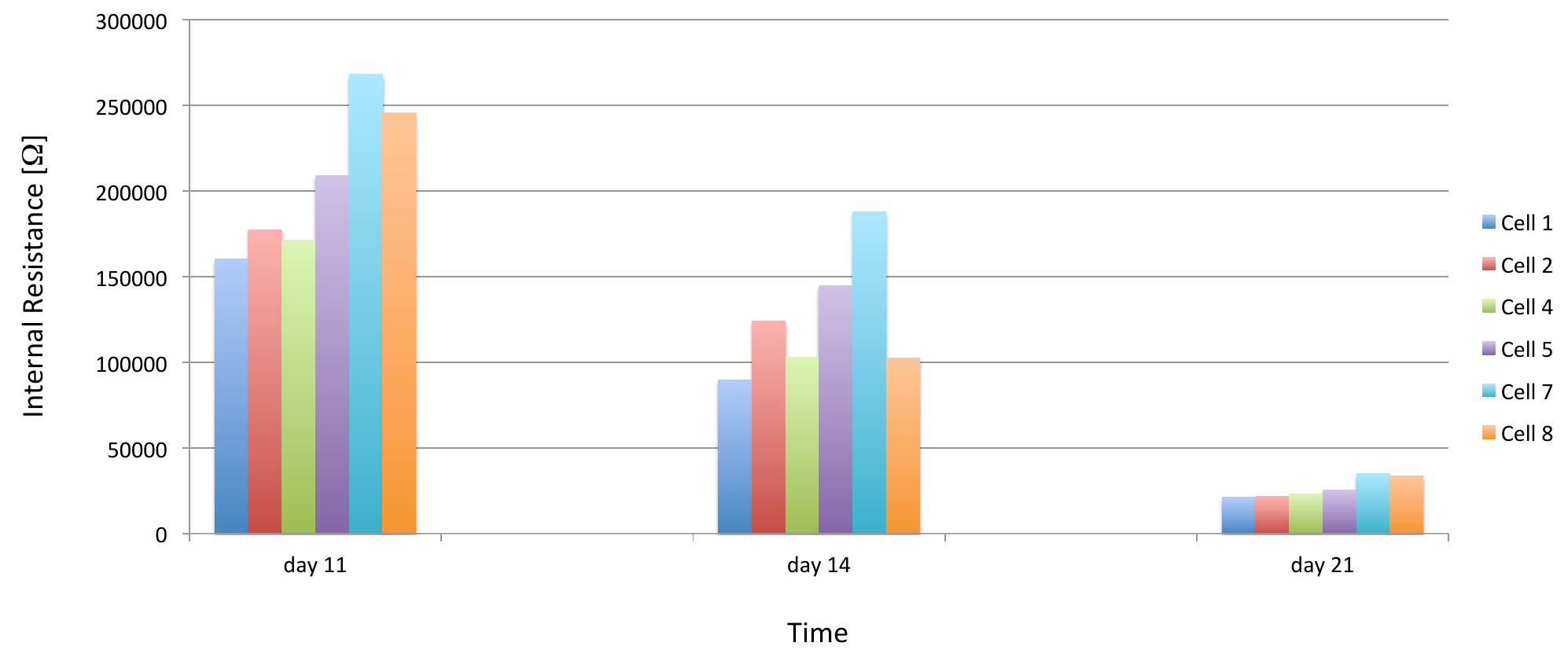}
	\caption{\label{Fig_int_res} Evolution of the internal resistance of the cells 1$\&$2,  4$\&$5 
	and 7$\&$8: the values of internal resistance and their evolution in time are very similar,
	 confirming the assessed reproducibility of our tests.}
\end{figure}

\subsection{Effect of salt water amendment}

\begin{figure}
\centering
	\subfigure[]{\includegraphics[angle=-90,width=0.49\textwidth]{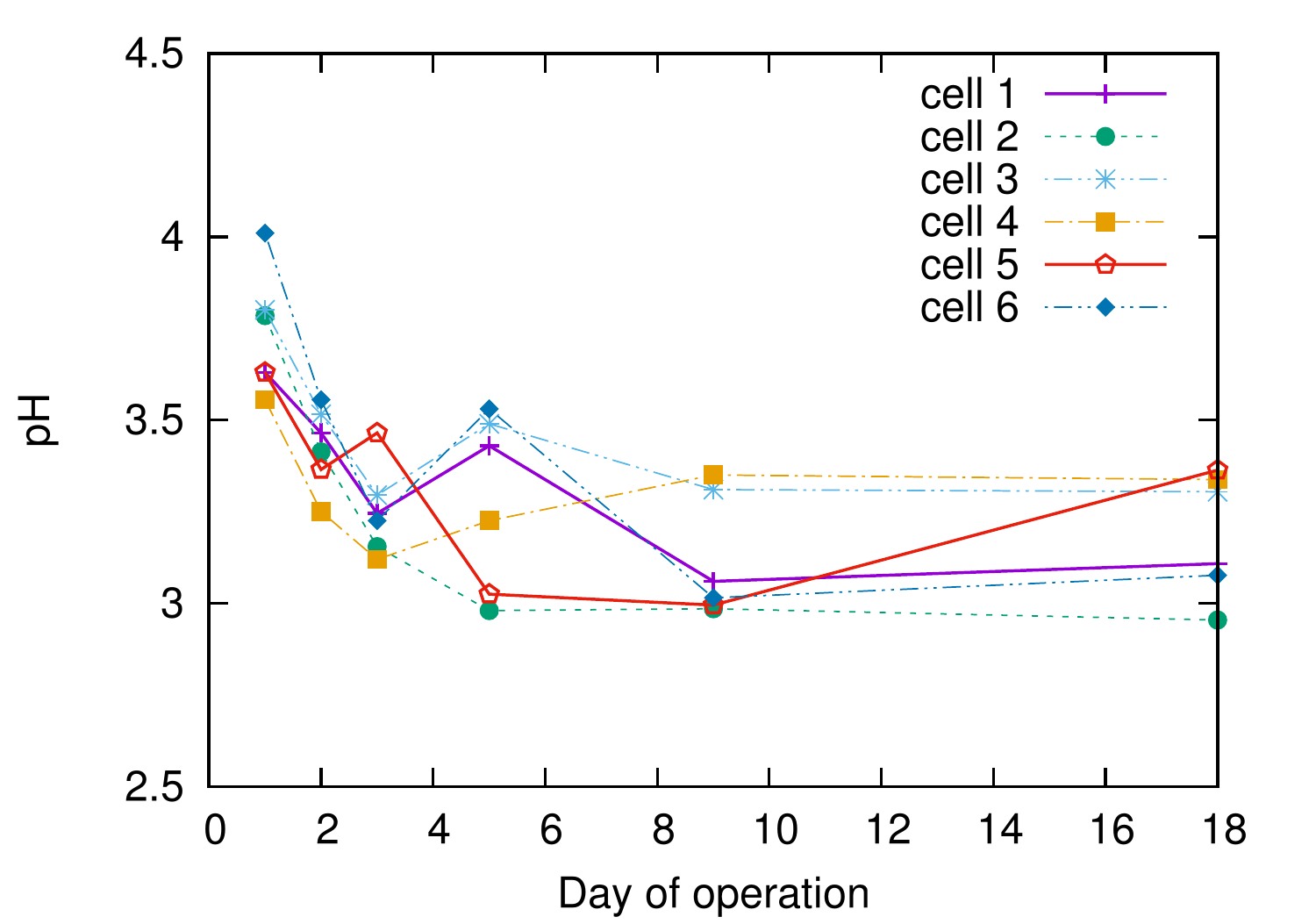}} \
	\subfigure[]{\includegraphics[angle=-90,width=0.49\textwidth]{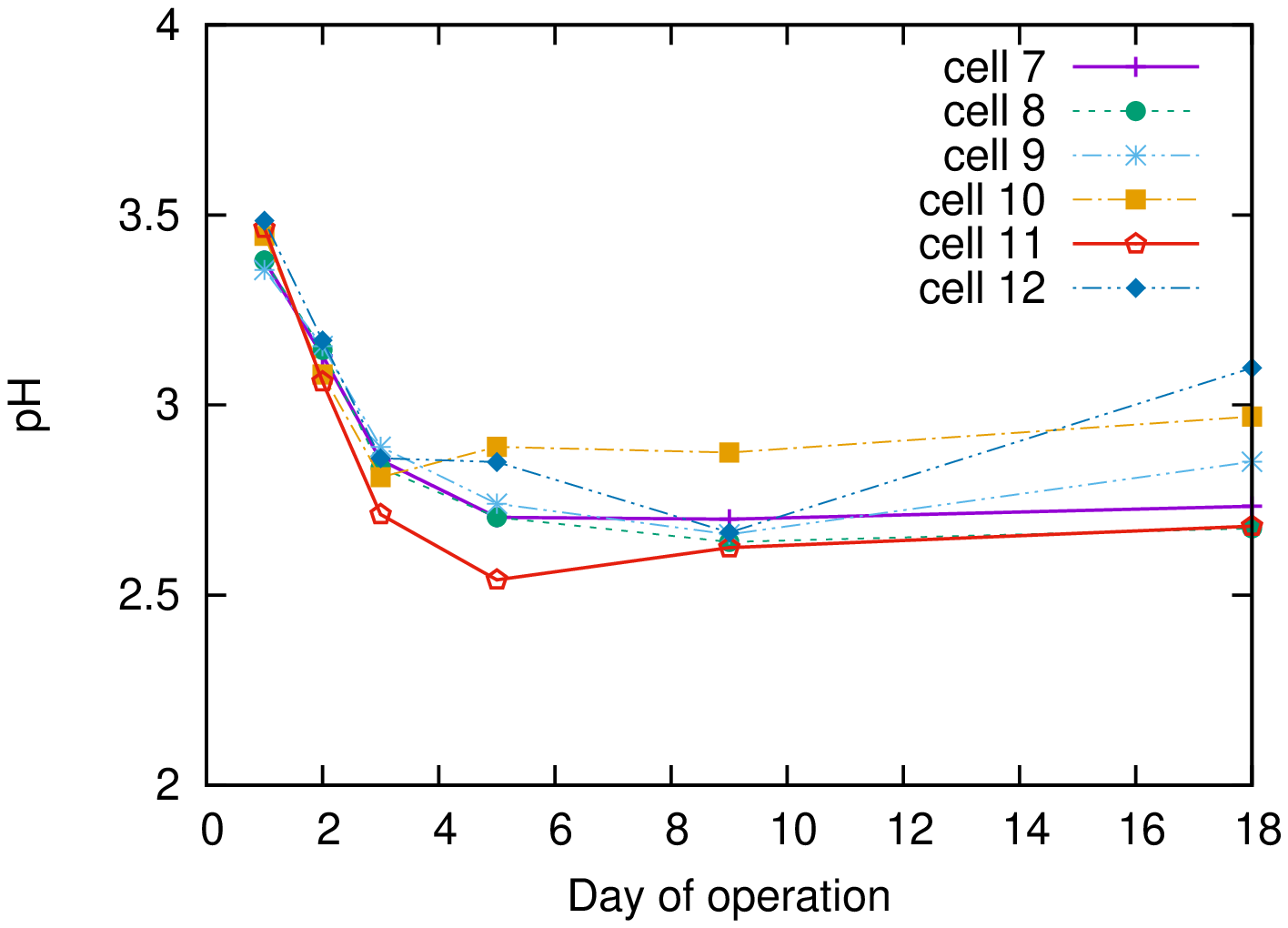}}
	\caption{\label{Fig_pH} Time evolution of feedstock pH inside our TMFC reactors: (a) reports the trend of fresh-water-based reactors, while in (b) the evolution of salt-water cell pH is highlighted.} 
\end{figure}

Figure \ref{Fig_pH} reports the trends of internal pH value for fresh-water cells (left panel) and for salt-water reactors (right panel). 
The salt-water TMFCs were characterized by a higher internal acidity, due to the membrane fouling. Nafion, in fact, has negatively-charged chemical groups which attract the positive ions in the organic substrate. Since only H$^+$ can pass through the membrane, the other positive ions accumulate at the cathode, preventing the further passage of protons: this causes the lower pH values shown in Fig. \ref{Fig_pH}(b).

\begin{figure*}
\centering
\subfigure[]{\includegraphics[angle=-90, width=0.49\textwidth]{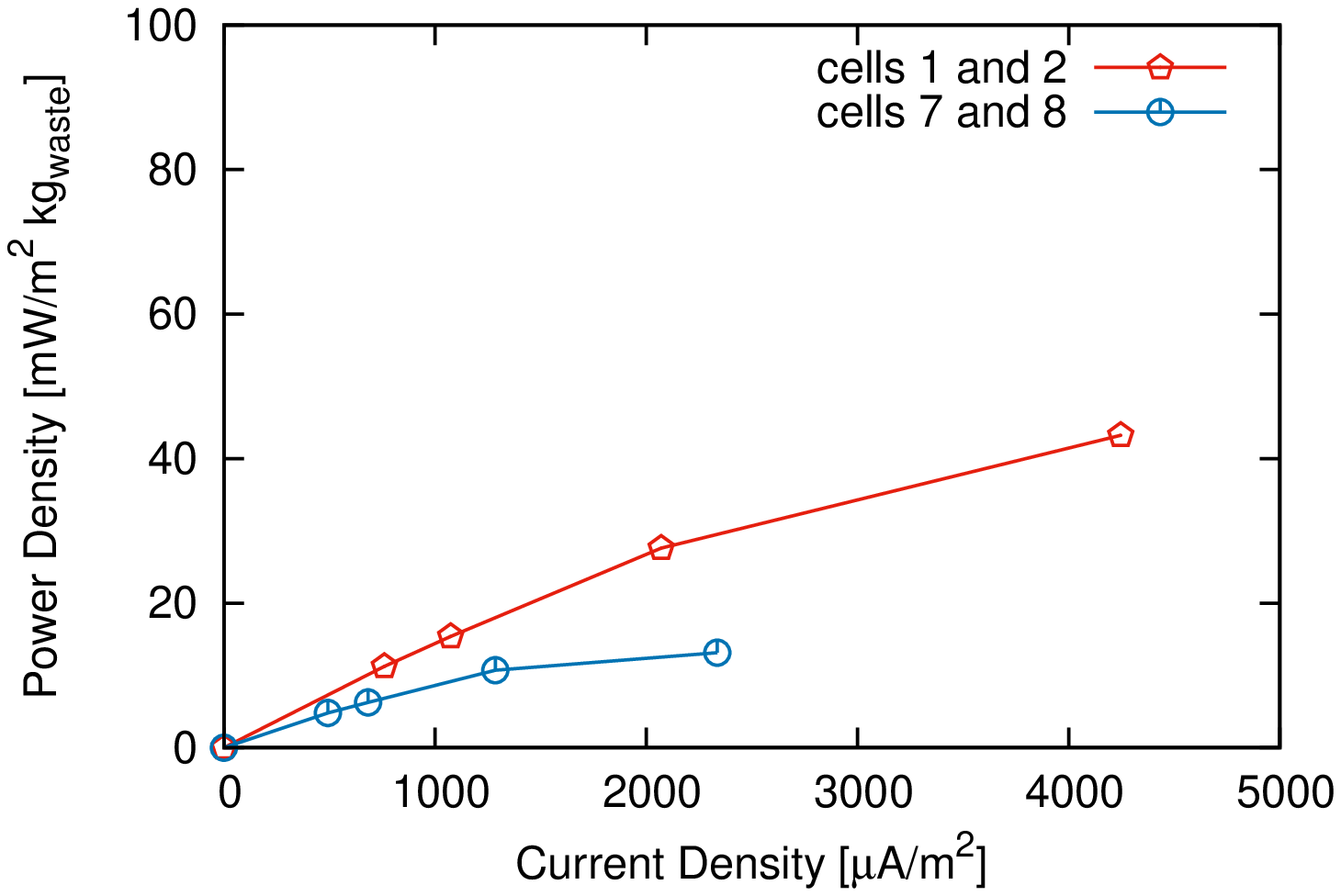}} \
\subfigure[]{\includegraphics[angle=-90, width=0.49\textwidth]{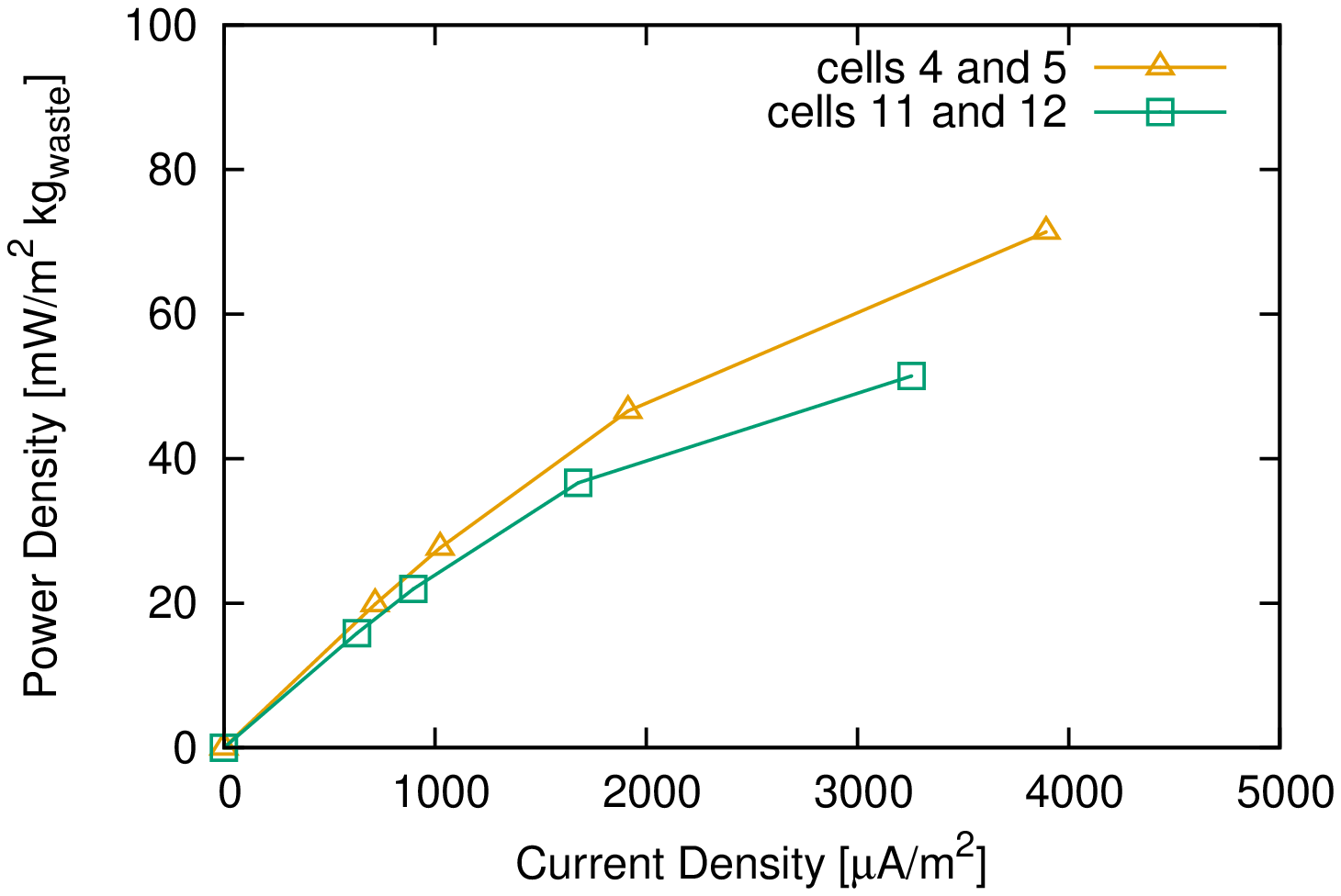}}
\caption{\label{fresh-salt} Comparison of MFC performances with fresh and salt-water.}
\end{figure*}

The effect of water composition on TMFC performance is reported in Figure \ref{fresh-salt}(a) for cells 1 and 2 vs cells 7 and 8 at day 14 and in Fig. \ref{fresh-salt}(b) for cells 4 and 5 vs 11 and 12. In both cases, the cells with NaCl addition provide lower performance in terms of power production.

\subsection{Effect of Potentiostatic Growth}

\begin{figure*}
\centering
\subfigure[]{\includegraphics[angle=-90,width=0.49\textwidth]{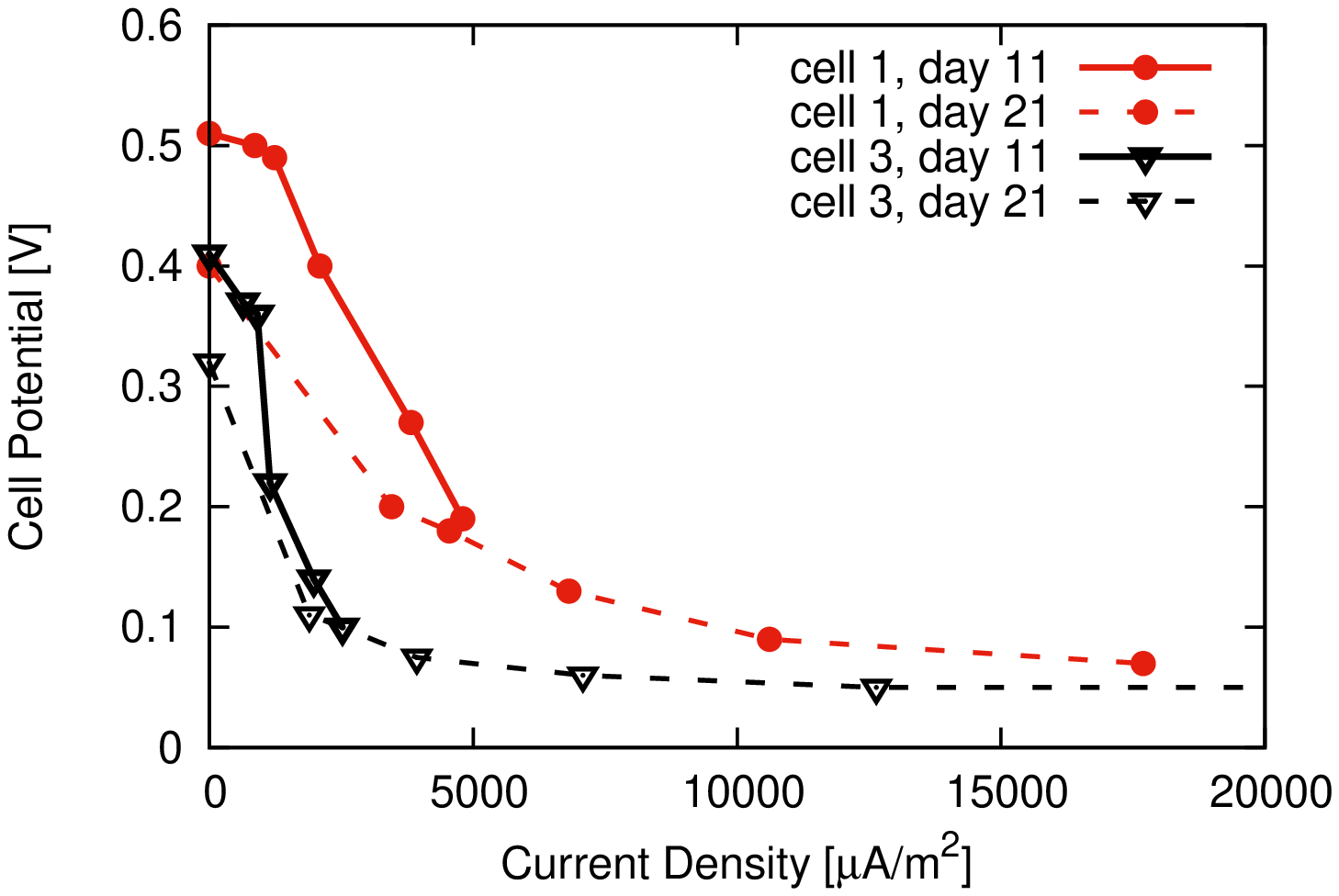}} \
\subfigure[]{\includegraphics[angle=-90,width=0.49\textwidth]{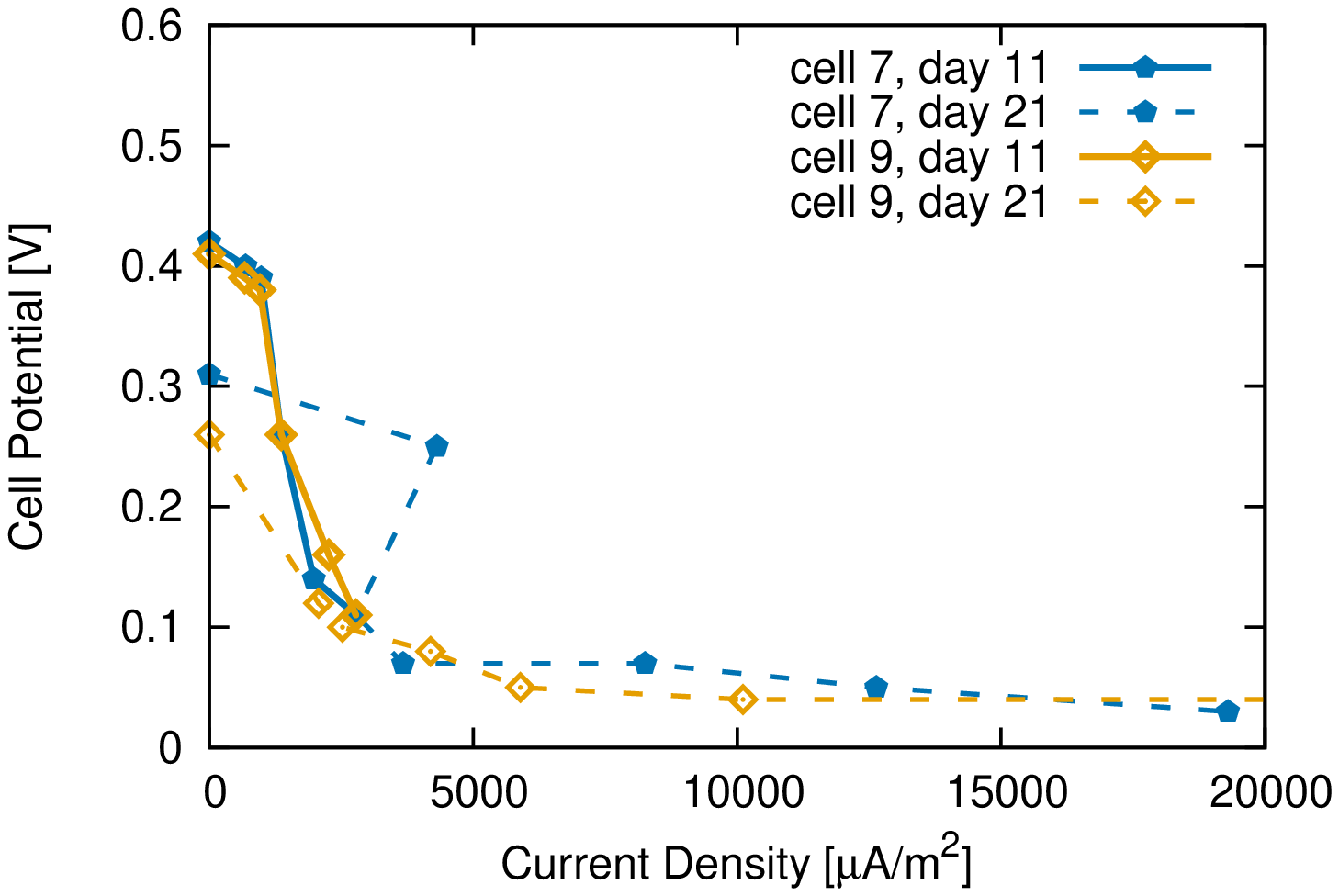}}
\caption{\label{pot_growth} Comparison of the performances of the MFC with and without potentiostatic growth at different  days of operation: (a) cell 1 vs cell 3; (b) cell 7 vs 9. The potenstiostatic growth does not show any enhancement in MFC performance.}
\end{figure*}

In Figure \ref{pot_growth}, the performance of TMFCs that were subject to a potentiostatic growth phase in the first week of operation are compared to the behavior of similar reactors (according to Tab. \ref{Tab_2}). \\
From this Figure, it is evident that carrying out the potentiostatic growth at the beginning of the experiment does not bring sensible enhancements in the power production of the reactors along the experiment, both for fresh and salt water systems.

According to the results in the literature, the potentiostatic growth is known to have a positive effect on the performance of MFCs operating with complex substrates, such as wastewaters with metal pollutants, \cite{Huang_2011_MFC}. 
In our case, on the contrary, it has a detrimental effect on the cell performance: this can be ascribed to the different characteristics of our substrate, which is rich in carbohydrates, sugar and proteins (typical with fruit and vegetable slurries), and to the presence of the Nafion membrane at the cathode.

\subsection{Effect of spent feedstock inoculum}

\begin{figure*}[h!]
\centering
\subfigure[]{\includegraphics[angle=-90,width=0.49\textwidth]{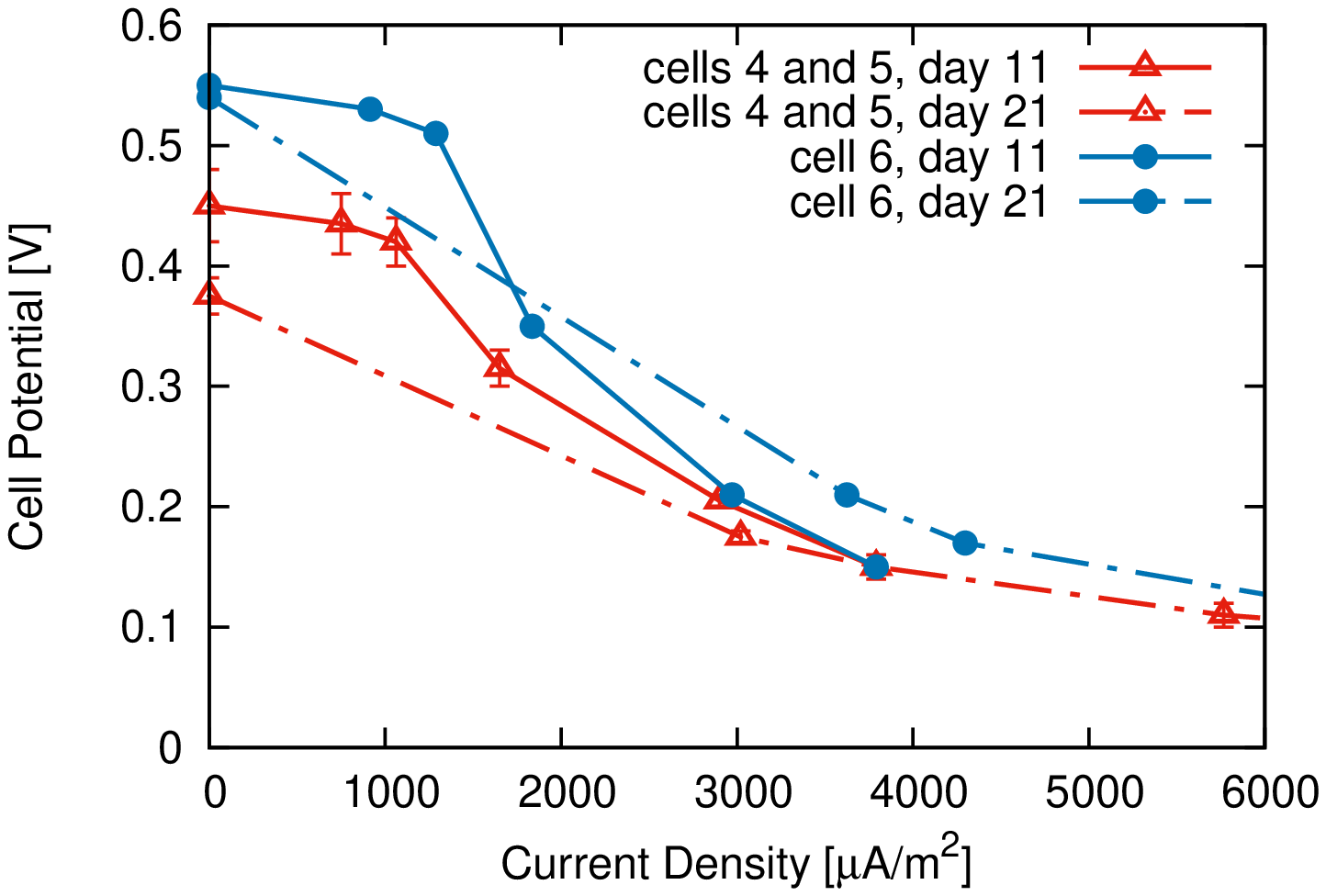}} \
\subfigure[]{\includegraphics[angle=-90,width=0.49\textwidth]{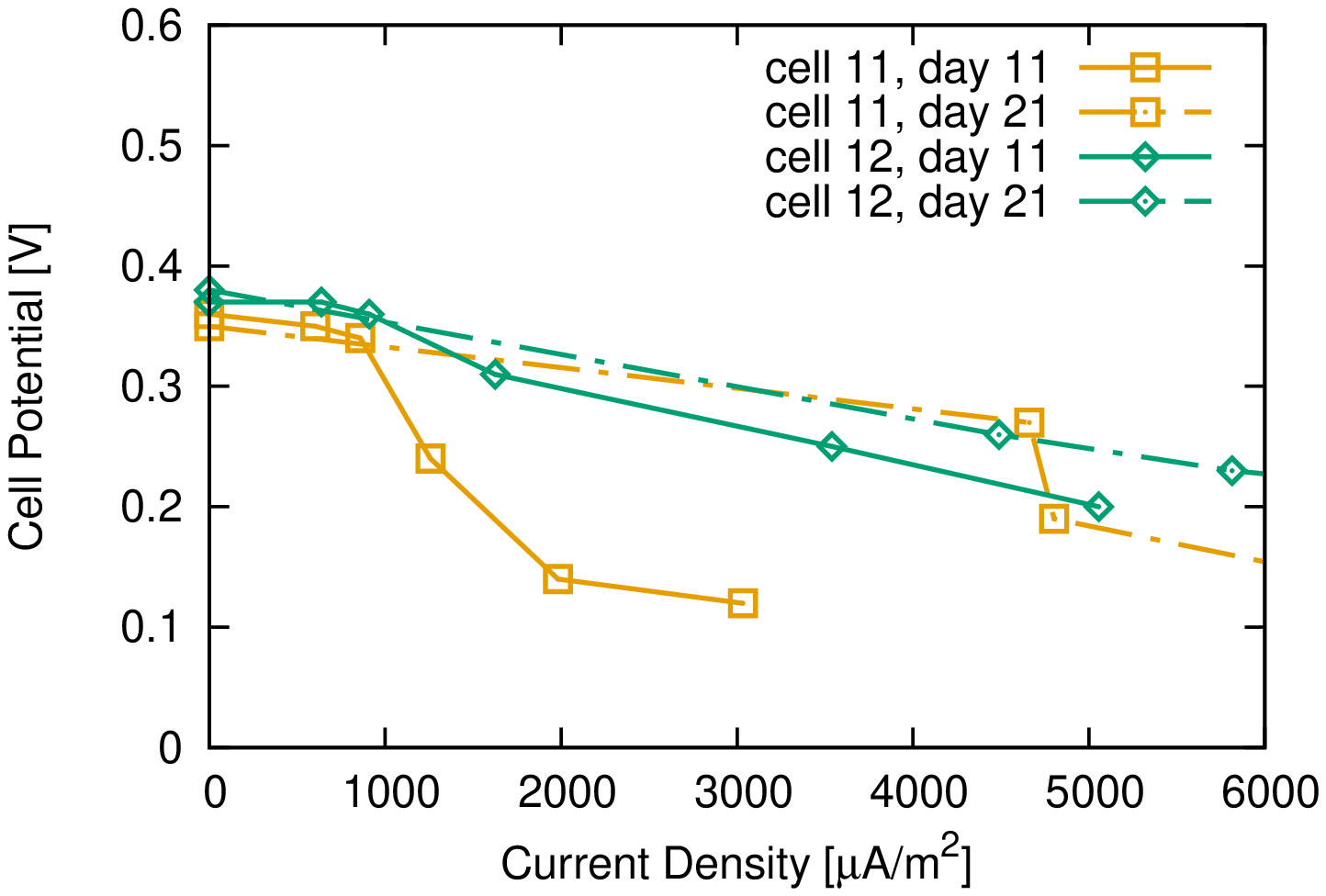}} \\
\subfigure[]{\includegraphics[angle=-90,width=0.49\textwidth]{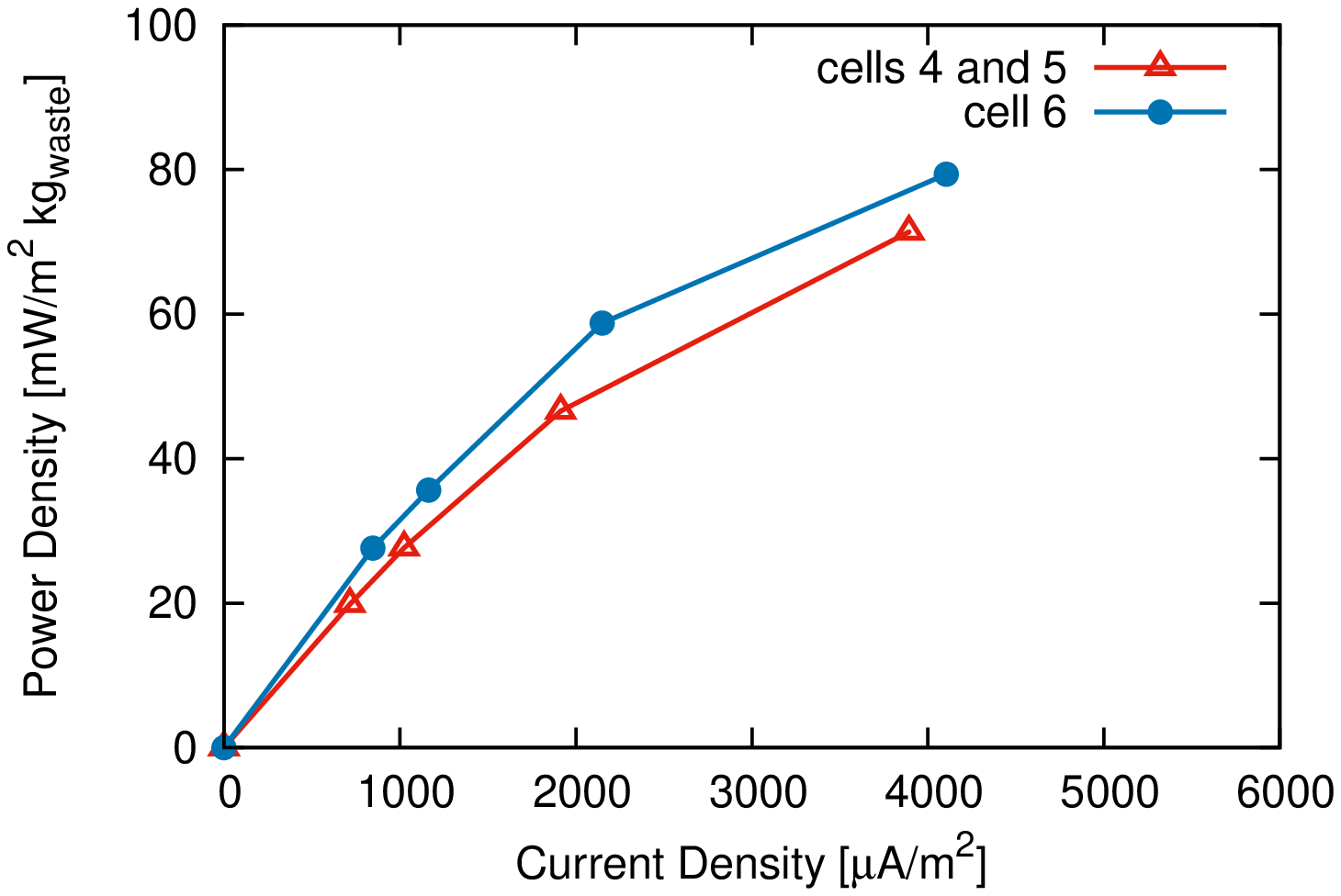}} \
\subfigure[]{\includegraphics[angle=-90,width=0.49\textwidth]{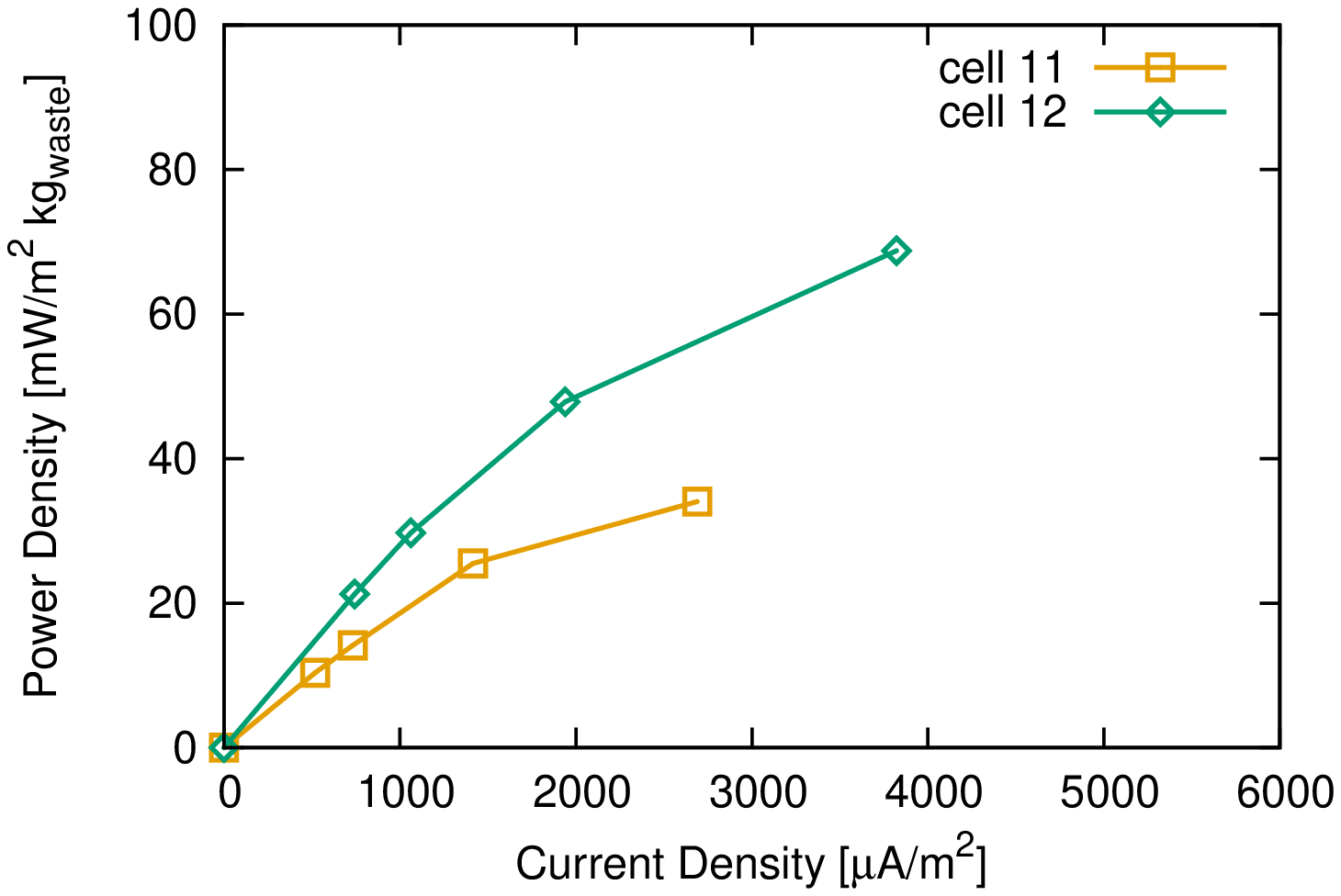}}
\caption{\label{inoc} Comparison of MFC performances with and without the inoculum of previously elaborated OFMSW: (a) and (b) report the polarization curves at different days of the experiment; (c) and (d) the power curves at day 14.}
\end{figure*}

Cells 4, 5 and 10, 11 have been prepared with the same water as 1, 2, 3 and 7, 8, 9 respectively, according to the details reported in the previous Sections. \\
The aim of inoculating pre-digested feedstock was to investigate whether an enriched microflora could increase the power production of TMFCs. \\
Figure \ref{inoc} shows the comparison of cell performance in terms of power production at different days of cells 4 and 5 vs 6 and 11 vs 12. Cell 10 has not been included, as its performances were too different from all other TMFCs. According to Fig. \ref{inoc}, the addition of pre-digested organic matter provides a performance decrease of $\sim 10 \%$ for freshwater reactors. In the case of salt-water TMFCs, the cell with the inculum has provided almost half the power production as the corresponding reactor with fresh solid waste. Nevertheless, a certain reduction of the cell activation energy was detected, as highlighted by the slope of the polarization curves at low current densities in Fig. \ref{inoc}.

Such a complex behavior in presence of a spent feedstock amendment can be ascribed to the very low pH values of the pre-digested substrate (pH~$\sim 2.5$) which contributes to further increasing the overall acidity inside the reactors, making the environmental conditions even harsher. \\
In the salty reactors, the increase in salt concentration leads to the further decrease in performance showed in Fig. \ref{inoc}: further studies are in progress to shed light on these outcomes.

\subsection{Effect of water dilution}

\begin{figure*}
\centering
\subfigure[]{\includegraphics[angle=-90,width=0.49\textwidth]{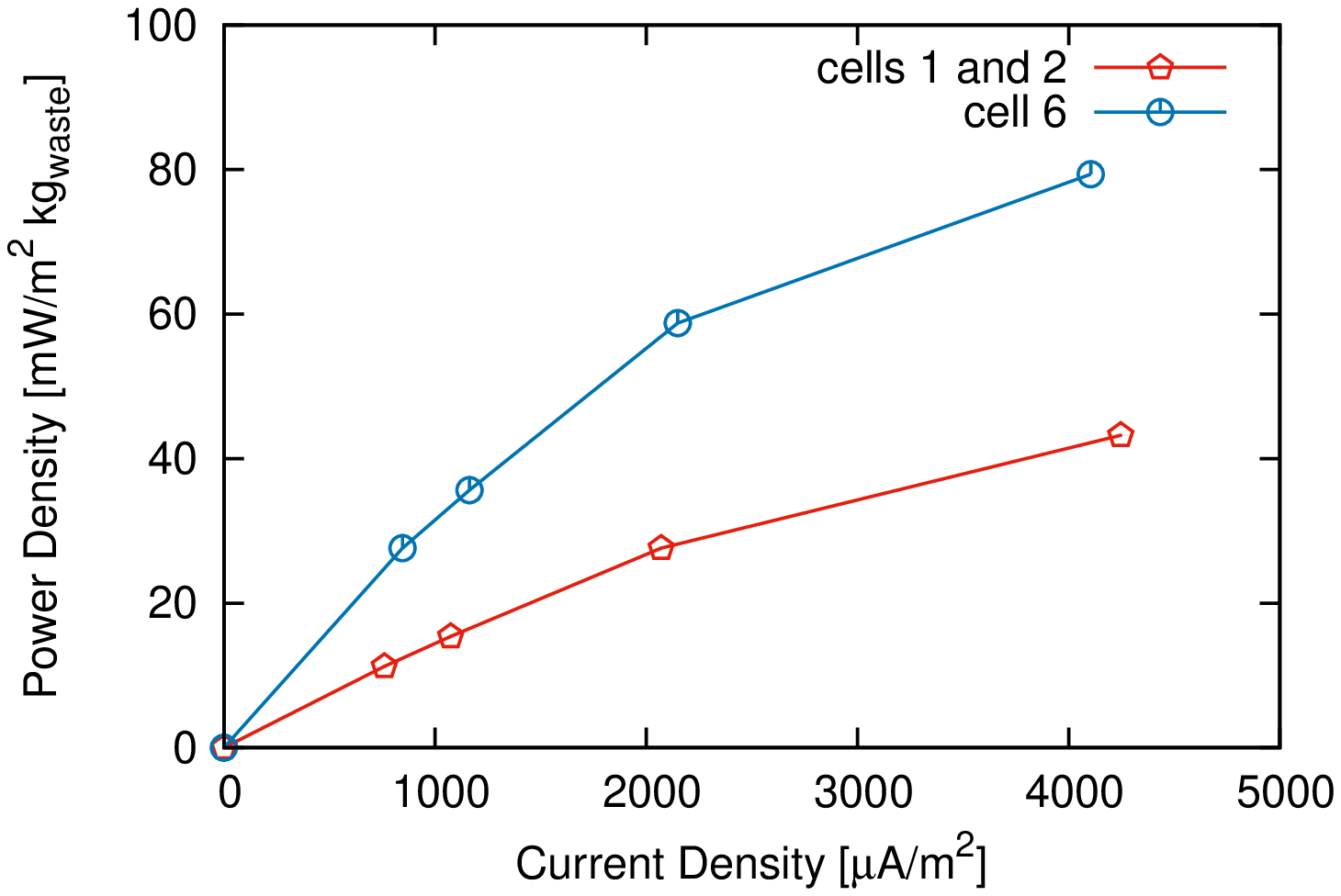}} \
\subfigure[]{\includegraphics[angle=-90,width=0.49\textwidth]{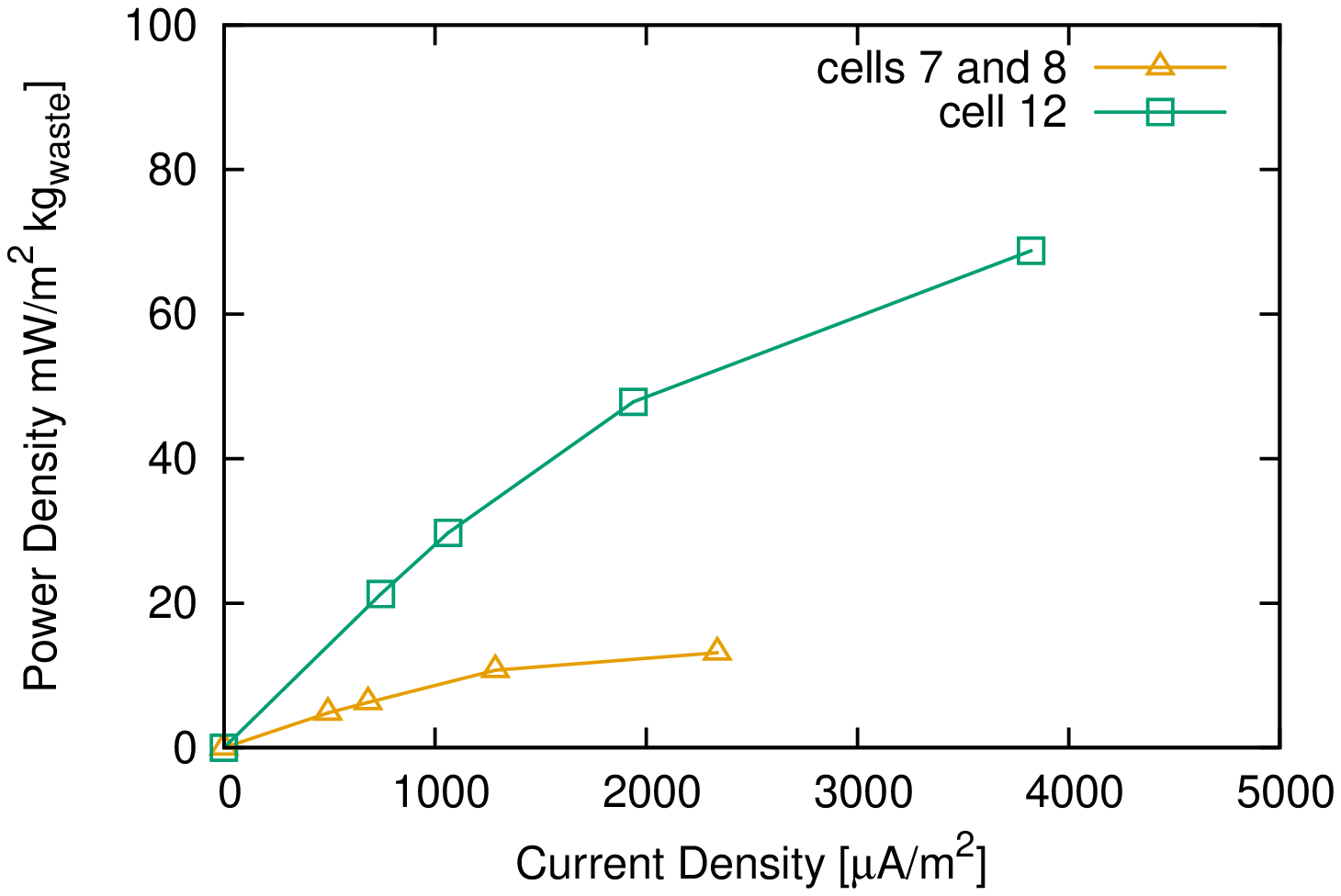}} 
\caption{\label{water_dil} Comparison of MFC performances with different values of water dilution.}
\end{figure*}

Figure \ref{water_dil} shows the comparison between power curves at the 14th day of operation for cells characterized by 1:1 and 3:1 Liquid-to-Solid ratio, for fresh water (Fig. \ref{water_dil}(a)) and salt water (Fig. \ref{water_dil}(b)) reactors. \\
In the case of fresh water, the two trends at the 14th day of operation are very close, while in the case of salt-water reactors, the power trend in more diluted MFCs is considerably above that of 1:1 reactors.
It is worth noting that in both the fresh- and salt-water reactors, the higher the feedstock dilution, the higher the power density.

\section{Industrial outlook}

It is interesting to foresee an industrial application for energy generation from MFCs based on
solid organic waste. Very likely, low pH values negatively affected tubular MFCs performances, nevertheless a power production at  $\sim 40$ mW/m$\cdot$kg$_\text{waste}$ can be estimated: in all the cases, in fact, such a power density value is easily reached both with fresh and salt water, with the solid organic content ranging between 15~g and 30~g . We can, thus, focus on this power value to study the feasibility of the scale-up of such reactors for large power production applications. As a benchmark, we consider the energy produced at a modern anaerobic digestion plant, which is accredited of processing 33000 tons per year of solid waste, producing 
$\sim 1$ MW of power, for a global energy production of $\sim 8000$ MWh/year, \cite{Caivano}. If we consider the same amount of waste, our TMFCs would produce:
\begin{equation}
	40 \times 10^{-3} \ \text{W/m}^2 \cdot \text{kg} \; \times \; 33 \cdot 10^6 \ \text{kg}  \sim \; 1000 \ \text{kW/m}^2 \; , \nonumber
\end{equation}
which is comparable to a modern anaerobic digestion plant of the same capability. \\
It is well known that scaling up MFCs causes issues related to a considerable increase in internal resistance, but the realization of proper anodic and cathodic surfaces can limit this problem. Our research on an innovative cathodic material is aimed at achieving a sustainable and economically cheap solution to this issue.

\section{Conclusions}

In this work we have assessed the performance of air cathode, single-chambered, Tubular Microbial Fuel Cells (TMFCs) provided with Nafion membrane, according to different operating conditions. \\
All the reactors have worked for 28 days at ambient temperature ($T = 20 \pm 2^{\circ}C$) with feedstocks characterized by low values of pH (pH$=3.0 \pm 0.5$) and high NaCl content (35 mg/L in 6 of the 12 reactors). \\
Our TMFCs proved to be capable of producing power even with such harsh internal conditions, showing a higher power yield in the cells with fresh water: the salty feedstock, in fact, is characterized by considerably low values of pH, probably due to the progressive deterioration and fouling of Nafion membrane.
For the industrial scaling-up of our reactors, then, the Nafion membrane will not be adopted both for its deterioration and for its prohibitive costs.\\
The potentiostatic growth at the beginning has provided some slight increase in power production only in presence of salt water and only for the first half of the experiment: thus, such a procedure is not considered as a valid tool to enhance MFC power productivity. \\
The inoculum of 10$\%$ of a solid organic waste previously digested for a 28-day period in an 
analogous experiment has not shown a positive effect on MFC performance, due to the low pH value reached at the end of the digestion period (pH $\sim 2.5$). Such a procedure, which is standard in anaerobic digestion plants for the development of methanogenic bacteria, is not recommended to improve MFC performance working in low-pH environments.

Finally, we have developed and employed a novel porous material at the cathode, which has shown promising performances and could be adopted in scaled-up MFC plants; nevertheless, further studies are needed to better characterize the graphite-ceramic material and to improve its chemical and mechanical properties.

The obtained results provide a further evidence of the versatility of MFC technology.

\section{Acknowledgments}

This work was supported by  the  
Italian Government Research Project PON03PE\_00109\_1  ``FCLab - Sistemi innovativi e tecnologie ad alta efficienza per la poligenerazione'', with Prof. Elio Jannelli as the Scientific Responsible. \\
The precious contributions of Dr. Giovanni Erme and Dr. Enzo De Santis for reactor realization are kindly acknowledged.


%

\end{document}